\documentclass[aps,prl,reprint]{revtex4-2}
\usepackage{graphicx}
\usepackage{bm}

\begin{document}


\title{Strongly localized wrinkling modes of single- and few-layer graphene sheets in or on a compliant matrix under compression} 



\author{Yu. A. Kosevich}
\email[]{Corresponding author: yukosevich@gmail.com}
\affiliation{N.N. Semenov Federal Research Center for Chemical Physics of Russian Academy of Sciences, 4 Kosygin Str., Moscow 119991, Russia,}
\affiliation{Plekhanov Russian University of Economics, Stremyanny per. 36., 117997 Moscow, Russia}

\author{I. A. Strelnikov}
\affiliation{N.N. Semenov Federal Research Center for Chemical Physics of Russian Academy of Sciences, 4 Kosygin Str., Moscow 119991, Russia,}

\date{\today}

\begin{abstract}
We present a study of the wrinkling modes, localized in the plane of single- and few-layer graphene sheets embedded in or placed on a compliant compressively strained matrix. We provide the analytical model based on nonlinear elasticity of the graphene sheet, which shows that the compressive surface stress results in spatial localization of the extended sinusoidal wrinkling mode with soliton-like envelope with localization length, decreasing with the overcritical external strain.
The parameters of the extended sinusoidal wrinkling modes are found from the conditions of anomalous softening of flexural surface acoustic wave propagating along the graphene sheet in or on the matrix. For relatively small overcritical external strain, the continuous transition occurs from the sinusoidal wrinkling modes with soliton-like envelope to the strongly localized modes with approximately one-period sinusoidal profiles and  amplitude- and external-strain-independent spatial widths. Two types of graphene wrinkling modes with different symmetry are described, when the in-plane antisymmetric or symmetric modes are presumably realized in the graphene sheet embedded in or placed on a compliant strained matrix. Strongly localized wrinkling modes can be realized without delamination of the graphene sheet from the compliant matrix and are not equivalent to the ripplocations in layered solids. Molecular-dynamics modeling confirms the appearance of sinusoidal wrinkling modes in single- and few-layer graphene sheets embedded in polyethylene  matrix at $T=300 K$.
\end{abstract}


\maketitle 


Exceptional physical and mechanical properties of graphene have made it very attractive for the construction of nano- and electro-mechanical devices and as a reinforcing inclusion in polymer nanocomposites \cite{geim07,lee08,akinwande17}.  As reinforcing layers in polymer nanocomposites, the graphene sheets provide longitudinal stiffness that significantly exceeds the corresponding characteristics of the polymer matrix, which ensures high tensile strength of the nanocomposite in the plane of reinforcement. However, under the compression conditions, the ultimate load of the nanocomposite is determined not by the strength of its components
but by the loss of stability of the stiff reinforcing layers embedded in the polymer matrix.

In the main approximation, the instability of a graphene sheet embedded in a compliant compressively strained matrix results in the appearance of extended sinusoidal wrinkling mode. Such bending instability was first described within the linear macroscopic theory of elasticity \cite{biot37,biot57,allen69}, and with the use of the Winkler model \cite{androulidakis14,androulidakis18} and molecular dynamics method \cite{lin15,koukar16} later. In this paper, based on nonlinear elasticity we show that that the compressive surface stress in the graphene sheet embedded in or placed on a compliant compressively strained matrix results in spatial localization of the extended sinusoidal wrinkling mode with soliton-like envelope with localization length, decreasing with the overcritical external strain.  We also show that for relatively small overcritical strain, the continuous transition occurs from the sinusoidal wrinkling modes with soliton-like envelope to the strongly localized modes with approximately one-period sinusoidal profiles and  amplitude- and surface-stress-independent spatial widths. The origin of the soliton-like envelope and localization of the sinusoidal wrinkling mode we relate with the effective attraction of the wrinkling modes in the nonlinear system. The parameters of the extended sinusoidal wrinkling mode are found from the conditions of anomalous softening of flexural surface acoustic wave (SAW) propagating along the two-dimensional elastic sheet placed in or on the soft matrix \cite{kos87a,kos87b,kos18a,kos18b}, which provide an analytical approach to get the necessary parameters of the model.

The main contribution to the elastic energy of the embedded graphene sheet are given by its in-plane strain $\epsilon_{\alpha\beta}$ produced by external uniform strain $\epsilon_{\alpha\beta}^0$, and out-of-plane  $w\equiv u_z^s$ and in-plane $u_{\alpha}^s$ displacements,
\begin{equation}
\epsilon_{\alpha\beta}=\epsilon_{\alpha\beta}^0+\frac{1}{2}(\frac{\partial u_{\alpha}^s}{\partial x_{\beta}}+\frac{\partial u_{\beta}^s}{\partial x_{\alpha}})+\frac{1}{2}\frac{\partial w}{\partial x_{\alpha}}\frac{\partial w}{\partial x_{\beta}},
\label{def}
\end{equation}
 when the total deformation energy of the two-dimensional (2D) elastic sheet (isotropic in its plane) has the following form:
 \begin{eqnarray}
E_{T}&=&\int\int[\frac{1}{2}\lambda_s\epsilon_{\alpha\alpha}^2+\mu_s\epsilon_{\alpha\beta}^2+\frac{1}{2}D_{11}\left (\frac{\partial^2 w}{\partial x^2}+\frac{\partial^2 w}{\partial y^2}\right )^2 \nonumber\\
&+&2D_{66}\left (\left (\frac{\partial^2w}{\partial x\partial y}\right )^2-\frac{\partial^2w}{\partial x^2}\frac{\partial^2w}{\partial y^2}\right ) \nonumber \\
&+&\frac{1}{2}K_B(k_1,k_2)w^2]dxdy,
\label{en}
 \end{eqnarray}
where $\alpha,\beta=1,2$, $\lambda_s$ and $\mu_s$ are the 2D $Lam\acute{e}$ coefficients, which determine the 2D elastic modulus tensor $h_{\alpha\beta\gamma\delta}$, $D_{11}$ and $D_{66}$ are the diagonal and torsional bending rigidities of the 2D elastic sheet, $K_B(k_1,k_2)>0$ is the positive coefficient, which describes the coupling of the out-of-plane static displacement $w(x_1,x_2)$ of the sheet with the matrix (or substrate), assuming the continuity of $w(x_1,x_2)$ and corresponding change of surface-projected bulk
stress $\sigma_{zz}$ at the sheet surface, see Eqs. (11) and (12) below.
Coefficient $K_B(k_1,k_2)$ depends only on the matrix (or substrate) bulk moduli of elasticity and 2D wave vectors $k_{1,2}$ of the sheet out-of-plane displacement $w(x_1,x_2)$ and is not related with the corresponding coefficient in the Winkler model \cite{androulidakis14}, see Eqs. (20)-(23) below. Essentially all the introduced 2D moduli of elasticity $\lambda_s$, $\mu_s$, $D_{11}$ and $D_{66}$ are determined by the in-plane (valence-bonds and valence-angles) interatomic interactions in the sheet and are finite even in the sheet with a monolayer thickness, see, e.g., \cite{kos97,kos18a,kos18b}. Equation (2) implies that nonlinear elasticity of the graphene sheet embedded in a compliant matrix is determined by the 2D $Lam\acute{e}$ coefficients $\lambda_s$ and $\mu_s$, which substantially  exceed the corresponding coefficients of the matrix (multiplied by interatomic distance).

To demonstrate the main features of the considered phenomenon, we consider the simplest case of one-component external compression in the x-direction, $\epsilon_{\alpha\beta}^0=\epsilon_{xx}^0<0$, when the problem reduces to the one-dimensional one with $w=w(x)$. Our main assumption, confirmed by the final results, is the displacement $w$ in the form of a static sinusoidal mode, long on the lattice-period scale, with the spatially-dependent amplitude, $w=A(x)\sin(k_0x)$,
 which describes, as we will show, a $soliton$-$like$ $envelope$ of the sinusoidal wrinkling mode. In this case, the coefficient $K_B(k_1,k_2)$ in (2) takes the form $K_B=Bk_0$, $k_0>0$ (proposed for the first time in Ref. \cite{biot37}), and the effective elastic modulus $B$ can be
determined from the spectrum of soft flexural SAW  in the system \cite{kos87a,kos87b,kos18a,kos18b}.  Substituting this ansatz for $w$ in Eq. (2) and averaging the sinusoidal-mode functions in the weakly-modulated limit $\mid A'\mid\ll k_0A$, we get the following deformation energy per unit area of the 2D sheet:
\begin{eqnarray}
E_{def}&=&E_0+\frac{1}{4}g_{xx}[A^2k_0^2+A^{'2}]+\frac{1}{4}D_s[A^2k_0^4+A^{''2}\nonumber \\
&+&4k_0^2A^{'2}-2k_0^2AA'']+ \frac{3}{64}E_s[A^4k_0^4+A^{'4}+2k_0^2A^2A^{'2}] \nonumber \\
&+&\frac{1}{4}Bk_0A^2,
\label{ent}
\end{eqnarray}
where $E_s=\lambda_s+2\mu_s$ is 2D Young modulus, $g_{xx}=E_s\epsilon_{xx}^0<0$ is the external-strain-induced surface stress, $D_s=D_{11}$, $E_0$ is surface energy of the flat graphene sheet.

The equilibrium form $A(x)$ and wrinkle wave number $k_0$ can be determined from the extremum conditions of surface energy $E_{def}$ with respect to $A$, $\delta E_{def}/\delta A=0$, and $k_0$, $\partial E_{def}/\partial k_0=0$. The latter partial derivative should be taken under the condition of homogeneity of the structure, $A^{'}=A^{''}=0$, because $k_0$ is by definition the characteristics of homogeneously wrinkled elastic sheet, which weakly depends on the amplitude $A$ and overcritical compression, see below. From the extremum conditions be obtain two equations,
which in the limit of weak modulation have the following form:
\begin{eqnarray}
\label{eqeq1}
D_sk_0^3=&\frac{1}{2}B,\\
-\frac{3}{8}E_sk_0^2A^3&=&(g_{xx}+3D_sk_0^2)A+(g_{xx}+6D_sk_0^2)A^{''}.
\label{eqeq2}
\end{eqnarray}
These equations give us the wave number of the wrinkled  structure $k_0=(B/2D_s)^{1/3}$ and the critical
in-plane surface stress and strain: $g_{xx}^{(cr.)}=E_s\epsilon_{xx}^{(cr.)}=-3D_sk_0^2$. Then Eq. (\ref{eqeq2}) takes the form of the Ginzburg-Landau-type equation for the external-strain-driven inhomogeneous scalar order parameter $A$ but with the opposite sign of the dispersive term, given by the coefficient in front of $A^{''}$:
\begin{eqnarray}
\frac{3}{8}k_0^4A^3&-&(\mid\epsilon_{xx}^{0}\mid-\mid\epsilon_{xx}^{(cr.)}\mid)k_0^2A \nonumber\\
&+&(2\mid\epsilon_{xx}^{(cr.)}\mid-\mid\epsilon_{xx}^{0}\mid)A^{''}=0.
\label{gl}
\end{eqnarray}
 For weakly overcritical strain, $2\mid\epsilon_{xx}^{(cr.)}\mid >\mid\epsilon_{xx}^{0}\mid\geq\mid\epsilon_{xx}^{(cr.)}\mid$, Eq. (\ref{gl}) describes the wrinkling sinusoidal modes with broken in-plane translational symmetry
 with soliton-like envelope:
 \begin{eqnarray}
 \label{env1}
 w_{s,a}&=&F\frac{(\cos(k_0x),\sin(k_0x))}{\cosh(\gamma x)}, \\
 \label{env2}
 F&=&\frac{4}{k_0\surd3}\sqrt{\mid\epsilon_{xx}^{0}\mid-\mid\epsilon_{xx}^{(cr.)}\mid}, \\
 \label{env3}
 \gamma&=&k_0\sqrt{\frac{\mid\epsilon_{xx}^{0}\mid-\mid\epsilon_{xx}^{(cr.)}\mid}{2\mid\epsilon_{xx}^{(cr.)}\mid-\mid\epsilon_{xx}^{0}\mid}}\ll k_0,
\end{eqnarray}
where $w_s$ and $w_a$ describe the symmetric and antisymmetric in the graphene plane wrinkling modes.

The origin of the soliton-like envelope of the wrinkling sinusoidal mode we relate with the combination of the repulsive nonlinearity of the embedded graphene sheet, given by the positive term $(3/64)E_Sk_0^4A^4$ in Eq. (\ref{ent}), and the $negative$ $effective$ $mass$ (NEM) of the soft flexural SAW,
given by the negative second derivative $\partial^2\omega/\partial k_x^2$ for the SAW at $k_x\simeq k_0$ beyond the softening, see Fig. 1.
These two features of the considered system result in the effective attraction of the wrinkling modes.
The appearance of the soliton-like envelope of the extended sinusoidal wrinkling mode is similar to the appearance of the envelope solitons and 
intrinsic localized vibrational modes (discrete breathers)
in the Fermi-Pasta-Ulam lattice with repulsive quartic nonlinearity \cite{kos74,siev88,page90,kos93a,flach98}, which can be related with the negative effective mass of short-wavelength acoustic phonons in the lattice and can appear in result of modulational instability of the lattice band-edge mode \cite{kos00,kos17}, and which can be of the symmetric or antisymmetric type (with even or odd parity) \cite{siev88,page90,kos93a}. The linear dispersive and nonlinear terms in the nonlinear envelope-function equation for short-wavelength excitations in hard anharmonic lattice \cite{kos93a} have the same signs as that in Eqs. (\ref{eqeq2}) and (\ref{gl}).

\begin{figure}
\includegraphics[width=7cm, height=5cm]{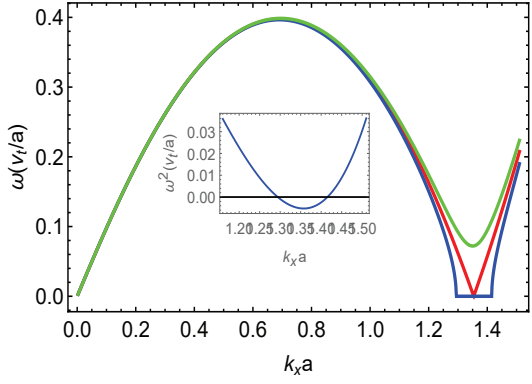}
\caption{\label{fig1} Anomalous softening of flexural SAW, propagating along the graphene monolayer embedded in a matrix
of polyethylene, caused by the compressive strain in
the propagation direction. Frequency $\omega$ is measured in units of
$v_t/a$, wave vector $k_x$ is measured in units of $1/a$, where $v_t = 0.861$
km/s, $a$=$\rho_s/\rho$= 7.6 {\AA} for the polyethylene with density $\rho$=998
kg/m$^3$. Green, red and blue lines correspond to the strain
0.998$\epsilon_{xx}^{(cr.)}$, 1.0$\epsilon_{xx}^{(cr.)}$ and 1.002$\epsilon_{xx}^{(cr.)}$, respectively. The blue line shows NEM of the SAW beyond the softening. Inset shows $\omega^2$ for the blue line close to $k_{x0}=1.35/a$, where $\omega^2<0$ and sinusoidal wrinkles with $k_x=k_{x0}$ exponentially grow in time. }
 \end{figure}

As follows from Eqs. (\ref{env1})-(\ref{env3}), amplitude $F$ of the broken-symmetry mode increases while the localization length $1/\gamma$ of the soliton-like envelope decreases with the increase of the overcritical external strain $\mid$$\epsilon_{xx}^{0}$$\mid$. For the large enough $\mid$$\epsilon_{xx}^{0}$$\mid$, but still less than  $2\mid$$\epsilon_{xx}^{(cr.)}$$\mid$, when $\gamma\sim k_0$, the modes with soliton-like envelopes (\ref{env1}) are replaced by strongly localized modes  with approximately one-period sinusoidal profiles with amplitude- and strain-independent spatial widths, cf. Ref. \cite{kos93a,kos93b,kos04}. For the symmetric and antisymmetric strongly localized modes,
we assume the following profiles:
\begin{equation}
\label{amode}
w_{s,a}=F^{(s,a)}(\cos(k_0x),\sin(k_0x))\cos^2(k_0x/2) 
\label{amode}
\end{equation}
for $-\pi/k_0<x<\pi/k_0$, and $w=0$ in the rest of the graphene sheet.  Substituting this ansatz in Eq. (2) and assuming the same form of the coefficient $K_B=\tilde{B}k_0$, we find the deformation surface energy $E_{def}^{(s,a)}$ as a function of $F$ and $k_0$. From the extremum conditions of  $E_{def}^{(s,a)}$ with respect to $F$ and $k_0$, we obtain the $k_0^{(s,a)}=\kappa^{(s,a)}(\tilde{B}/D_s)^{1/3}$, $\kappa^{(s)}=0.56$, $\kappa^{(a)}=0.52$, the critical strain $\epsilon_{xx}^{(cr.s,a)}=\varepsilon^{(s,a)}D_s^{1/3}\tilde{B}^{2/3}/E_s$, $\varepsilon^{(s)}=2.35$, $\varepsilon^{(a)}=1.80$, and the expression for the localized modes amplitudes $F^{(s,a)}=(1.89/k_0)\sqrt{\mid\epsilon_{xx}^{0}\mid -\mid\epsilon_{xx}^{(cr.s,a)}\mid}$.
In the crude assumption of $\tilde{B}=B$, we see that $\epsilon_{xx}^{(cr.s)}=1.24\epsilon_{xx}^{(cr. small)}$, $\epsilon_{xx}^{(cr.a)}=0.95\epsilon_{xx}^{(cr. small)}$, where $\epsilon_{xx}^{(cr. small)}$ is determined in the small-strain limit, described by Eqs. (4) and (5). But taking into account that $\tilde{B}>B$ because of the presence of higher spatial harmonics in the profiles (10), we conclude that the transition from the soliton-like envelope to the strongly localized wrinkling modes in the graphene sheet occurs continuously for relatively small overcritical strains, with $\mid$$\epsilon_{xx}^{(cr. small)}$$\mid<\mid$$\epsilon_{xx}^{(cr.s,a)}$$\mid<2\mid$$\epsilon_{xx}^{(cr. small)}$$\mid$, for which Eqs. (6)-(9) are also valid.
This is also clear from the comparison of the profiles of soliton-like solutions (7) with relatively large $\gamma\sim k_0$ with that of strongly localized solutions (10), see Fig. 2.

\begin{figure}
\includegraphics[width=7cm, height=5cm]{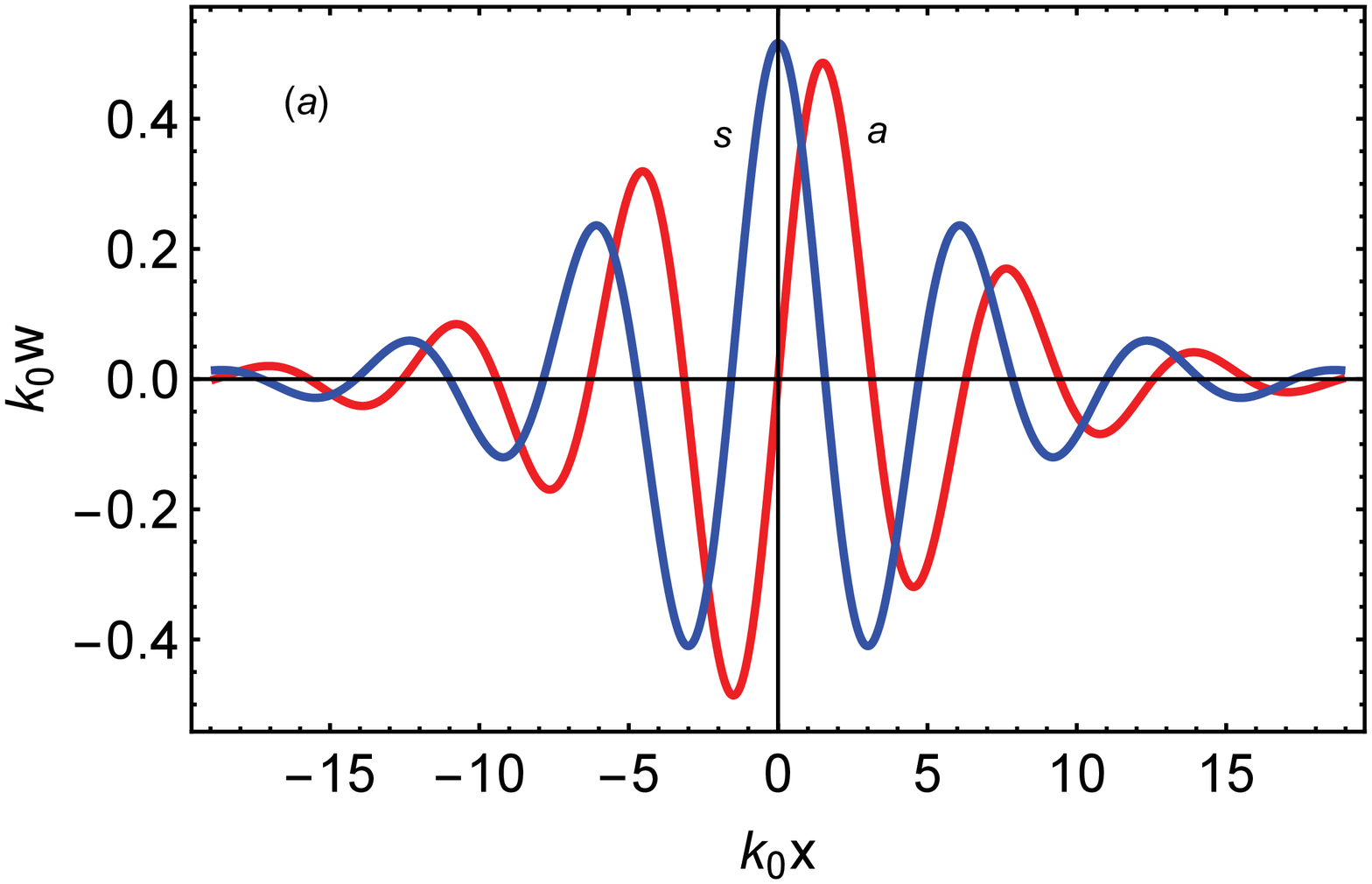}
\includegraphics[width=7cm, height=5cm]{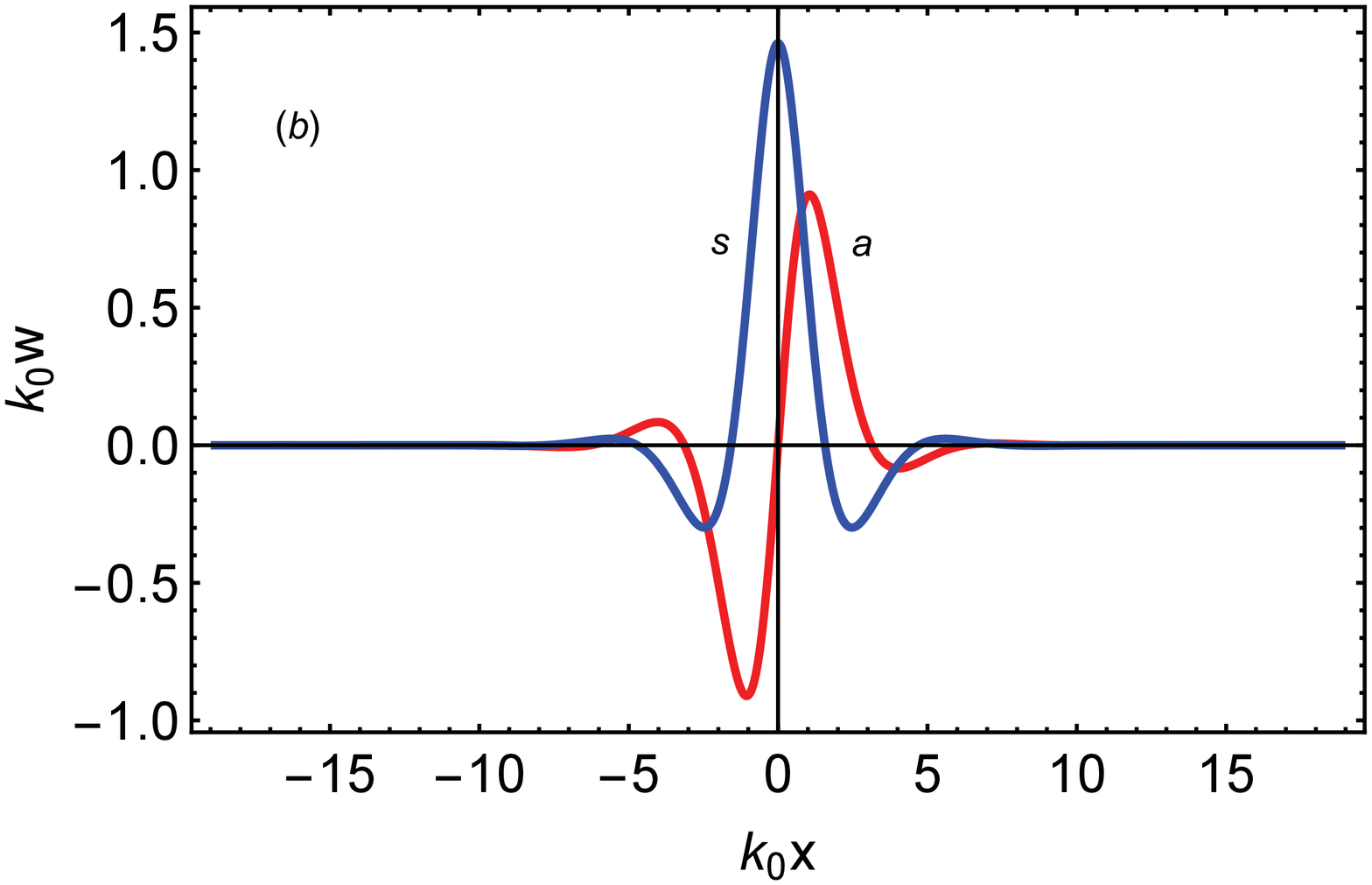}
\includegraphics[width=7cm, height=5cm]{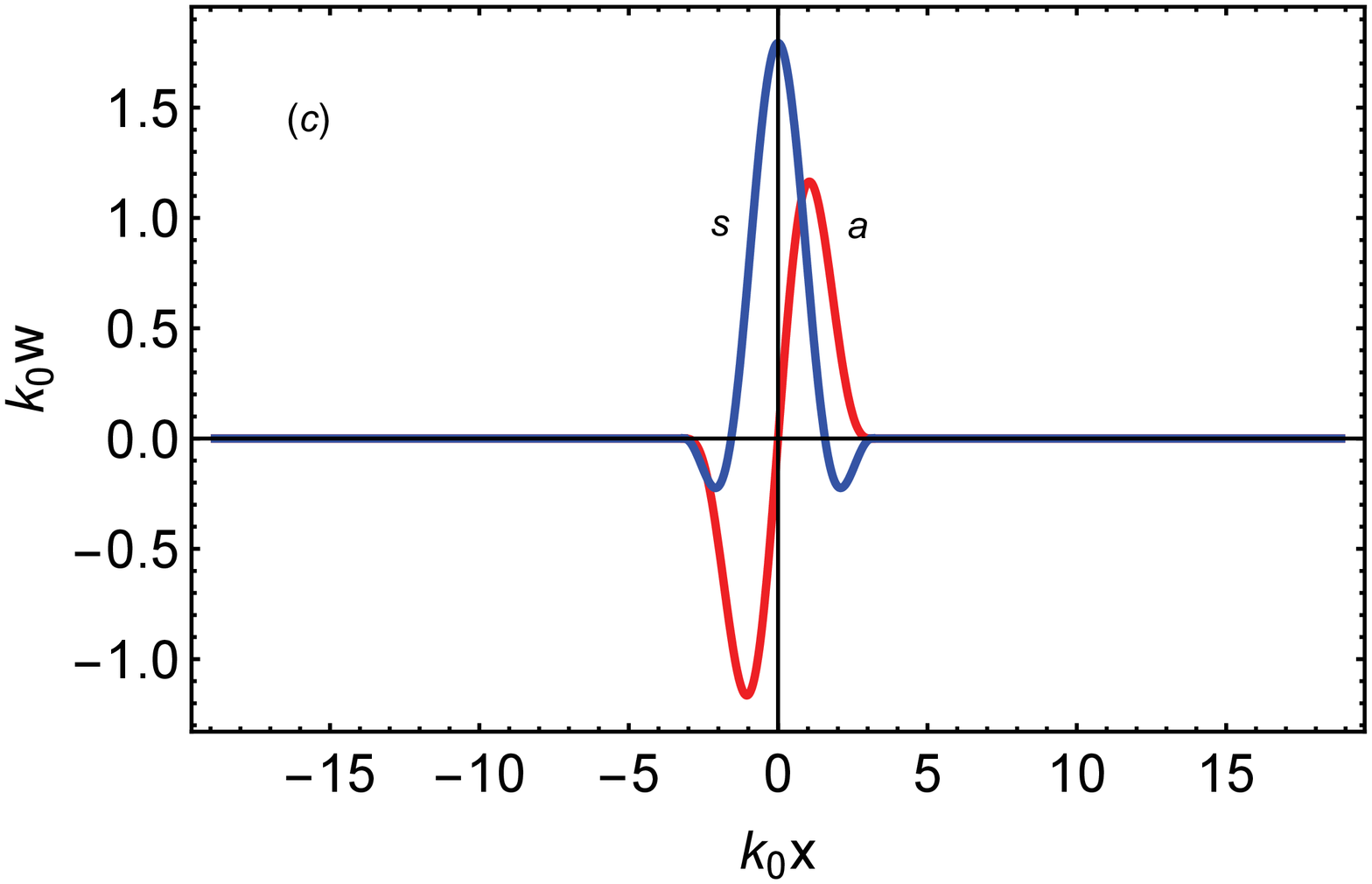}
\caption{\label{fig1} Reduced profiles of symmetric ``s" and antisymmetric ``a" wrinkling modes with soliton-like envelopes (7), for (a) $\mid\epsilon_{xx}^{0}\mid$=1.05$\mid\epsilon_{xx}^{(cr.)}\mid$ and (b) $\mid\epsilon_{xx}^{0}\mid$=1.4$\mid\epsilon_{xx}^{(cr.)}\mid$, and of strongly localized modes (10) for (c) $\mid\epsilon_{xx}^{0}\mid$=1.9$\mid\epsilon_{xx}^{(cr.)}\mid$.}
 \end{figure}

We should underline that the symmetric solutions (7) and (10) break while the antisymmetric solutions (7) and (10) do not break the {\it up-down symmetry} of the wrinkling mode: the sign of the integral $\int w_sdx\propto F$ along the sheet length is determined by the sign of $F$ and does not depend on $k_0$ and $\gamma$, while the integral is zero for the antisymmetric modes, $\int w_adx=0$. Therefore the antisymmetric modes are presumably realized in the graphene sheets embedded in the bulk of a compliant matrix when the up-down symmetry is clearly conserved in the defect-free system, while the symmetric modes are realized in the graphene sheets, placed on a compliant substrate (at the crystal-vacuum interface). In the latter case, the up-down symmetry in the wrinkling modes is approximately conserved for weakly overcritical compression of the graphene sheet, when $\gamma\ll k_0$, see Fig. 2(a).  Essentially the strongly localized wrinkling modes can be realized without delamination of the graphene from the compliant matrix or substrate. This feature can be related with the smallness of the graphene sheet bending rigidity $D_s$ in conjunction with its extremely high 2D Young modulus $E_s$. Because of the absence of delamination, the strongly localized
wrinkling modes in the graphene sheet embedded in a compliant matrix are not equivalent to the ripplocations in layered solids \cite{kush15,bars20} although there is an effective attraction between the localized
wrinkling modes similar to that of ripplocations \cite{bars20}. 

To find an expression for the coefficient $K_B=Bk_0$ in Eq. (2) in the case of anisotropic elastic matrix, we turn to the description of the softening of flexural surface acoustic wave (SAW) in the system, cf. \cite{kos18a,kos18b} for the case of isotropic matrix. The long-wavelength dynamical properties of the graphene sheet can be taken into account
with the use of dynamic boundary conditions for the displacements $u_i^{(1,2)}$ and surface-projected elastic stresses $\sigma_{ni}^{(1,2)}$ in the contacting bulk media in the vicinity of the embedded 2D elastic layer, see, e.g., \cite{andreev,kos97}. Dynamic boundary conditions, that are consistent with deformation energy (2),  have the following form:
\begin{eqnarray}
u_i^{(1)}=u_i^{(2)}&\equiv&u_i^s,\label{bc1}\\
\sigma_{ni}^{(1)}-\sigma_{ni}^{(2)}&=&g_{\alpha\beta}\nabla_\alpha\nabla_\beta u_i^s+\delta_{i\beta}h_{\alpha\beta\gamma\delta}\nabla_\alpha u_{\gamma\delta}^s\nonumber \\
&-&\delta_{iz}D_s\triangle_\alpha^2u_i^s
-\rho_s\partial^2 u_i^{s}/\partial t^2,\label{bc2}
\end{eqnarray}
where $\sigma_{ni}=\sigma_{ik}n_k$, $n_k$ is a unit vector of the normal to the interface directed from the medium 1 into medium 2, $i$=$1,2,3$,   $\rho_s$ is 2D mass density of the graphene sheet.
  When graphene sheet is placed at the crystal-vacuum interface,
we assume the slipping contact between the stiff 2D elastic layer and compliant substrate, under which the in-plane displacements $u_{\alpha}^s$  can be neglected in Eqs. (12) and boundary conditions in the substrate reduce to
\begin{eqnarray}
\sigma_{nz}&=&g_{\alpha\beta}\nabla_\alpha\nabla_\beta w -D_s\triangle_\alpha^2 w -\rho_s\frac{\partial^2 w}{\partial t^2},\label{bc2}\\
\sigma_{n\alpha}&=0. \label{bc2}
\end{eqnarray}

In general, we assume the orthorhombic symmetry of the compliant matrix, which can be a polyethylene crystal \cite{strnkv19}. First we consider the graphene sheet placed in $(001)$, $z=0$,  plane and the SAW propagates in the $[100]$ direction. Using boundary conditions (11) and (12) and following the approach in Ref. \cite{kos91}, we introduce the two-component elastic displacements $u_{x,z}^{(1,2)}$ in both contacting media in the sagittal $xz$ plane assuming the symmetric and antisymmetric with respect to the graphene plane distribution of the $u_z$ and $u_x$ displacements, when $u_x^s(0)=0$, and get the dispersion equation:
\begin{eqnarray}
&&(\rho\omega^2-C_{44}k_x^2)(\gamma_1+\gamma_2)k_x C_{33}=\frac{1}{2}(\rho_s\omega^2 -\nonumber \\
&&g_{xx}k_x^2-D_sk_x^4)(\rho\omega^2-C_{44}k_x^2-C_{33}\gamma_1\gamma_2 k_x^2),
\label{pssw}
\end{eqnarray}
where the elastic moduli $C_{ik}$ of the matrix are written in the Voigt notation, $\omega$ is the wave frequency, $k_x$ is the wavenumber, $\gamma_{1,2}$ are the parameters (with positive real component) that determine the inverse penetration depths of the two displacement components, $u_{z,x}^{(1,2)}\propto\exp(ik_x x\pm\gamma_{1,2}k_x z-i\omega t)$, which can be found from the equation \cite{kos85}
\begin{eqnarray}
&&C_{33}C_{44}\gamma^4-\gamma^2[C_{33}(C_{11}-\rho\omega^2/k_x^2)+C_{44}(C_{44}-\rho\omega^2/k_x^2)- \nonumber\\
&&(C_{13}+C_{44})^2k_x^2]+(C_{11}-\rho\omega^2/k_x^2)(C_{44}-\rho\omega^2/k_x^2)=0.
\label{pssw}
\end{eqnarray}

Similarly, using boundary conditions (13) and (14) at the crystal-vacuum interface we get the dispersion equation for the
Rayleigh SAW propagating
in $[100]$ direction on $(001)$ surface of orthorhombic crystal:
\begin{eqnarray}
&&\sqrt{C_{44}k_x^2-\rho\omega^2}[(C_{11}C_{33}-C_{13}^2)k_x^2-C_{33}\rho\omega^2] \nonumber \\
&-&\rho\omega^2\sqrt{C_{33}C_{44}(C_{11}k_x^2-\rho\omega^2)}=(\rho_s\omega^2\\
&-&g_{xx}k_x^2-D_sk_x^4)\sqrt{C_{33}C_{44}(C_{11}k_x^2-\rho\omega^2)}(\gamma_1+\gamma_2)k_x. \nonumber
\end{eqnarray}

The value of the critical negative surface stress  $g_{xx}^{(cr.)}$=$E_{s}$$\epsilon_{xx}^{(cr.)}$
and the wavenumber $k_{x0}$ at which the anomalous softening of the SAW occurs can be found from the following two conditions, cf. Refs. \cite{kos87a,kos87b,kos18a,kos18b}:
\begin{equation}
\omega(k_{x0})=0, ~~~~ \frac{\partial\omega(k_{x0})}{\partial k_x}=0.
\label{softw}
\end{equation}

With the use of Eqs. (15)-(18), we find the wrinkle wavenumber $k_{x0}$ and critical compressive in-plane
strain $\epsilon_{xx}^{(cr.)}$=$-3D_sk_{x0}^2/E_s$:
\begin{eqnarray}
\label{softw1}
 &&k_{x0}=\left[\frac{B_{bulk,subs}}{2D_s}\right]^{1/3},\mid\epsilon_{xx}^{(cr.)}\mid=\frac{3D_s^{1/3}}{E_s}\left[\frac{B_{bulk,subs}}{2}\right]^{2/3},\\
 &&B_{bulk}=\frac{2C_{44}C_{33}G}{C_{44}+\sqrt{C_{11}C_{33}}},~~B_{subs}=\frac{C_{11}C_{33}-C_{13}^2}{G\sqrt{C_{11}C_{33}}},\\
 &&G=\sqrt{2\sqrt{\frac{C_{11}}{C_{33}}}+\frac{C_{11}}{C_{44}}+\frac{C_{44}}{C_{33}}-\frac{(C_{13}+C_{44})^2}{C_{33}C_{44}}},
\label{softw2}
\end{eqnarray}
where $B_{bulk}$ or $B_{subs}$ is the effective elastic modulus, which determines parameter $K_B$  in Eq. (2) for the graphene sheet embedded in or placed on a compliant matrix, respectively. For $\mid$$\epsilon_{xx}^0$$\mid$$>$$\mid$$\epsilon_{xx}^{(cr.)}$$\mid$, we have $\omega^2(k_{x0})=-\Gamma^2(k_{x0})<0$ and the sinusoidal wrinkles $w(k_x x,\Gamma t)$ grow with maximal growth rate $\Gamma(k_{x0})$ at $k_x=k_{x0}$, which determines the wrinkle wavenumber $k_{x0}$, see Fig. 1. During the exponential growth of $w$, until the saturation imposed by the nonlinearity in deformation energy (3),
the NEM of soft flexural SAW and effective attraction between the wrinkling modes come into the play.

Expressions (20) are simplified for the isotropic (or transversally isotropic in $xz$ plane) matrix, when $C_{11}$=$C_{33}$, $C_{11}-C_{13}$=$2C_{44}\equiv 2\mu$ and $G$=$\gamma_1+\gamma_2$=$2$:
\begin{eqnarray}
\label{softw1}
B_{bulk}&=&\frac{8\mu(1-\sigma)}{3-4\sigma}=\frac{4E_b(1-\sigma)}{(1+\sigma)(3-4\sigma)},\\
B_{subs}&=&\frac{\mu}{1-\sigma}=\frac{E_b}{2(1-\sigma^2)}.
\label{softw2}
\end{eqnarray}
where $\mu$, $E_b$ and $\sigma$ are shear and Young moduli and Poisson's ratio of the bulk matrix.

It is worth mentioning that the modulus $B_{subs}$, given by Eq. (23), coincides with that obtained in the pioneer paper \cite{biot37}, but the modulus $B_{bulk}$, given by Eq. (22), is different from that obtained in the paper \cite{biot57}, in which it was assumed the relation $B_{bulk}$=$2 B_{subs}$, which was also used in later papers, see, e.g., Ref. \cite{groen01}. This relation holds only in the limit $\sigma\rightarrow$$1/2$ when  $B_{bulk}\rightarrow$$2 B_{subs}\rightarrow$0. On the other hand,
our expression (22) for $B_{bulk}$ is fully consistent with the value of the displacement of an infinite isotropic elastic medium in the plane normal to the applied local force (bulk Green's tensor) \cite{ll07}.  Our approach, based on the softening of flexural SAW, is macroscopic and does not rely on the particular model of the interaction of the graphene sheet with the matrix, like the Winkler model, because only the joint out-of-plane $w$   (and negligible in-plane $u_{\alpha}^s$) displacements of the sheet and matrix govern the long-wave bending instability.

It is important to emphasize that the bending instability of a single- or few-layer graphene sheet embedded in a compliant matrix, described by Eqs. (19)-(21), can also be applied to the bending instability of multilayer nanocomposite made from graphene sheets embedded in compliant matrix with the period $d$, which exceeds double penetration depths of both static elastic modes, $d>>2/(\gamma_{1,2}k_{x0})$, Eq. (16). On the other hand, taking in Eqs. (15) and (16) for isotropic matrix the limit $C_{44}k_x^2$$\ll$$\rho\omega^2$$\ll$$C_{11}k_x^2$ we obtain the dispersion of the internal wave, localized near the 2D elastic layer in an inviscid liquid: $\omega^2=(g_{xx}k_x^3+D_{s}k_x^5)/(2\rho+\rho_{s}k_x)$. Therefore Eqs. (15), (16) and (18) can also be used for the study of the bending of a swimming elastic foil \cite{argentina07,feinberg07} or of an inner floating layer of viscous liquid under compression with an account for the layer inertia effects, when the effective dynamical bending stiffness of the layer is determined by its thickness $d$ and viscosity $\eta$, $D_s$$\rightarrow$$-i\omega\eta d^3/3$, and corresponding $g_{xx}$ is determined by the in-plane compressive stress and two-side surface tension \cite{howell96,oratis20}.

For the graphene sheet with $n$ layers, when both $D_s(n)$ and $E_s(n)$ increase with $n$, Eqs. (19)-(21) predict the decrease with $n$ of the wrinkle  wavenumber $k_{x0}$ and modulus of the critical in-plane compression strain $\mid$$\epsilon_{xx}^{(cr.)}$$\mid$ \cite{androulidakis18,kos18b}, that demonstrates the non-Eulerian nature of the bending instability of the 2D elastic layer placed on or embedded in a compliant matrix \cite{androulidakis18}. In the case of weak interlayer coupling in the few-layer graphene, when $D_s(n)=nD_s(1)$ and $E_s(n)=nE_s(1)$ \cite{zach10}, Eqs.  (19)-(21) predict $k_{x0}\propto n^{-1/3}$ and $\mid$$\epsilon_{xx}^{(cr.)}$$\mid$$\propto n^{-2/3}$, cf. \cite{androulidakis18,kos18b}.

Molecular-dynamics modeling at
$T=300 K$ confirms the appearance of sinusoidal  wrinkling modes in single- and few-layer graphene sheets embedded in polyethylene  matrix at the critical compressive strain, see Fig. 3. Orthorhombic polyethylene matrix, with macromolecular chains along the zigzag graphene direction ("y" axis in the considered geometry), is described with the use of the model of Ref. \cite{paul}, in which the  polyethylene crystal at 300 K is in a soft rotational phase, transversally isotropic in $xz$ plane,  with density 998 kg/m$^3$, shear modulus 0.74 GPa, and ratio of longitudinal and transverse sound velocities in $xz$ plane  $v_l/v_t$=2.97. The single- and few-layer graphene sheets are described within the model of Ref. \cite{savin10}, with $\rho_s$=7.6$\cdot$10$^{-7}$ kg/m$^2$, $D_s(1)$=2.352$\cdot 10^{-19}$ N$\cdot$m and $E_s(1)$=139.1 N/m.  The increase of the wrinkle wavelength $\lambda_{x0}=2\pi/k_{x0}$ with $n$ is clearly seen in Fig. 3, and quantitative agreement is reached between the predicted by Eqs. (19)-(21) values of $\lambda_{x0}$ for $n=1,2,3$ and that obtained in the 
modeling, see also Fig. 1.

\begin{figure}
\includegraphics[width=7cm, height=5cm]{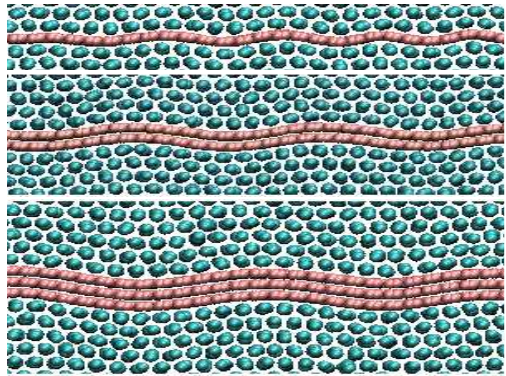}
\caption{\label{fig2} Visualization of polyethylene  matrix with sinusoidally wrinkled embedded single-, double- and triple-layer graphene sheet at weakly overcritical compressive strain at $T=300 K$,
from top to bottom.}
 \end{figure}

In conclusion, we provide the analytical model based on nonlinear elasticity of a
single- or few-layer graphene sheet, which shows that beyond the critical compressive surface stress the spatial localization of the wrinkling mode
occurs with soliton-like envelope with the localization length, decreasing with the overcritical compressive surface stress, that finally results in the formation of strongly localized mode with approximately one-period sinusoidal form and strain-independent length. Antisymmetric and symmetric wrinkling modes are described that presumably are realized in the graphene sheets embedded in or placed on a compliant strained matrix, respectively.

The authors thank A.V. Savin for useful discussions. This work was supported by the Russian Foundation for Basic Research, Grant No. 18-29-19135. This study was carried out
with the use of supercomputers of the Joint Supercomputer Center of Russian Academy of Sciences.

%
%

%



\end{document}